\documentclass[a4paper,11pt]{article}
\pdfoutput=1 

\usepackage{jinstpub} 

\graphicspath{ {./Images/} }    

\usepackage{multicol}           

\title{Fast high-power thyristors triggered in impact-ionization wave mode}

\author[a,b,1]{A.~Gusev,\note{Corresponding author.}}
\author[a]{S.~Lyubutin,}
\author[a,b]{V.~Patrakov,}
\author[a]{S.~Rukin,}
\author[a]{B.~Slovikovsky,}
\author[c]{M.~J.~Barnes,}
\author[c]{T.~Kramer,}
\author[c]{and V.~Senaj}

\affiliation[a]{Institute of Electrophysics UB RAS,\\106 Amundsen Street, Yekaterinburg, Russia}
\affiliation[b]{Ural Federal University,\\ 19 Mira Street, Yekaterinburg, Russia}
\affiliation[c]{CERN,\\CH-1211, Geneve 23, Switzerland}

\emailAdd{gusev@iep.uran.ru}

\abstract{GTO-like thyristors 5STH-2045H0002 (4.5~kV, 18~kA/\textmu s) developed by ABB semiconductors are currently used at CERN in LHC Beam Dumping System (LBDS): high-power switches with high dI/dt capability and low turn-on delay time are required. Implementation of the impact-ionization triggering in GTO-like thyristor enhances its switching performance and gives new information about semiconductor physics. In this work thyristors of 5STH-2045H0002 type triggered in impact-ionization wave mode are investigated. An SOS generator providing a dV/dt of several kV/ns was used as a source of triggering pulses. A thyristor switching time of approximately 200--300~ps was observed. Maximum discharge parameters were obtained for two series connected thyristors at a charging voltage of 10~kV, and a capacitor stored energy of ~300~J: peak current of 43~kA, dI/dt of 120~kA/\textmu s (limited by the discharge circuit), FWHM of 1.5~\textmu s. A single thyristor was tested in the repetitive mode at the charging voltage of 4.2~kV, and the stored energy of 18~J: peak current of 5.5~kA, dI/dt of 40~kA/\textmu s, FWHM of 1.5~\textmu s were obtained. No thyristor degradation was observed after more than one million pulses at a PRF up to 1~kHz in burst mode. Thyristor recovery time was 250~\textmu s. The switching efficiency was up to 98\% depending on dV/dt and stored energy.}


\begin{document}
\maketitle
\flushbottom

\section{Introduction}
\label{sec:intro}
Historically, the modulators of kicker systems at CERN typically use thyratrons as the high power switches and utilize either Pulse Forming Networks or Pulse Forming Lines (PFNs/PFLs)~\cite{Barnes_2018}. The highest voltage kicker systems in the CERN Proton Synchrotron (PS) use PFLs with $SF_6$ gas to enhance the insulation of the PFL. Due to environmental concerns as well as the difficulty of obtaining a suitable replacement for the $SF_6$ PFL, an alternative technology is required. In addition, thyratrons are susceptible to erratic (untriggered) turn-on, which cannot be accepted for future accelerators with significantly higher beam energies. Thus, solid-state technology is being studied for replacing thyratrons and PFLs in several systems at CERN, in particular in the Proton Synchrotron (PS)~\cite{Woog_2018}. The PS systems require magnet voltages of up to 40 kV, and trapezoidal current pulses with a rise time of approximately 30 ns and peak currents of approximately 3 kA.

A 100 TeV center-of-mass energy frontier proton collider, in a new tunnel of 80--100 km circumference, is a central part of CERN's Future Circular Colliders (FCC) design study \cite{FCC_2019}. The designs of the injection and extraction systems of the FCC are initially based upon the parameters of the injection and extraction systems of the Large Hadron Collider and a preliminary study of the FCC beam optics and lattice. The injection and, in particular, the extraction systems of the FCC have to be highly reliable. In order to achieve high-reliability, solid-state switches will be used for the generators of the injection and extraction systems.

The FCC extraction kicker system comprises the extraction kickers themselves as well as beam dilution kickers, both of which will be part of the FCC beam dump system and will have to reliably abort proton beams with stored energies in the range of up to 8.5 GJ. Both systems have to be precisely synchronized with the particle-free abort gap and have to track the beam energy from injection to flat top energy. An unsynchronized beam abort or particle in the abort gap would lead to a so-called "asynchronous beam abort". In this situation, particles are swept over the accelerator surface into the dump channel until the design trajectory is reached. Whilst in the LHC, sophisticated protection devices protect the accelerator components; an asynchronous dump event would be a catastrophic failure case for the FCC. The present specification for the FCC extraction kicker field rise time is less than 1~\textmu s. Due to eddy currents in the beam screen of the kicker magnet, the current produced by the solid-state generator must compensate by overshooting the nominal current. The requirements for the generator call for 5 kV and 4 kA output and a dI/dt is in the range of 10~kA/\textmu s \cite{FCC_2019}.

This work continues research aimed at replacing gas switches in pulsed power modulators by semiconductor devices. The phenomenon of sub-nanosecond current switching due to the delayed avalanche breakdown of Si p-n junctions was discovered by Grekhov et al.~in 1979 \cite{Grekhov_1979}. Later this switching mechanism was explained by initiation and propagation of an impact-ionization wave front in the semiconductor structure, and some semiconductor devices were developed based on this principle of operation \cite{Grekhov_2008,Grekhov_2010}. However, this effect was observed not only in specially designed devices but also in any Si devices with p-n junctions, including commonly used low-frequency thyristors \cite{Gusev_2015}.

Recently, silicon low dI/dt (0.4~kA/\textmu s) high-voltage (2.4~kV) thyristors produced by diffusion technology were investigated in impact ionization-wave mode for use as fast high-power switches for pulsed power applications \cite{Gusev_2016, Gusev_2017}. Switching in the impact-ionization wave mode occurs due to applying a steep overvoltage pulse across the thyristor's main electrodes --- anode and cathode: the gate electrode remains in the open-state. However, there are technological shunts in the thyristor structure; these shunts ensure a gate-cathode connection and have a resistance of about 0.6~$\Omega$. A voltage rate of rise, dV/dt, of the overvoltage pulse of more than 1~kV/ns is required. Power of the triggering generator has to be sufficient for increasing the voltage across the thyristor to approximately twice the static breakdown voltage. Under such conditions, the thyristors go into the conductive state within \verb'~'200 ps. Switches based on thyristors with a diameter of the silicon wafer of 32--56~mm are currently being designed, with the following range of the switching parameters: blocking voltage up to 20~kV, peak current up to 220~kA, maximum current rate of rise up to 130~kA/\textmu s, and FWHM up to 25~\textmu s.

However, there is a lack of research of thyristors with a highly interdigitated gate and cathode structure operated in impact-ionization wave mode. These thyristors have common anode and base layers, but the cathode layer is manufactured using integrated circuit technology. These thyristors can be considered as large devices comprising of a huge amount of small thyristors placed on a single chip. An integrated gate-commutated thyristor (IGCT), gate turn-off (GTO) thyristor or any GTO-like structure can be considered as an example of this. There is no commonly used name for this type of thyristor. Therefore, it is referred to here as a GTO-like thyristor. One should keep in mind that this name refers only to the structure, and not the triggering mode in our case. GTO-like thyristors were developed for pulsed power applications. They have blocking voltage of \verb'~'5 kV, maximum peak current up to 150~kA, and maximum current rate of rise, dI/dt, of up to 50~kA/\textmu s~\cite{Welleman_2009}. GTO-like thyristors can be used for the replacement of gas switches such as thyratrons and ignitrons \cite{Welleman_2001}. State of the art GTO-like devices have dI/dt capabilities above 100~kA/\textmu s \cite{Waldron_2015}.

The aim of the present work is to study the operation of the GTO-like thyristors triggered in impact-ionization wave mode. The main goal is to compare GTO-like thyristors with conventional thyristors operating in this mode. Switches triggered in impact-ionization wave mode can be a promising substitute for existing switches used in high-power accelerator facilities. One of the applications could be the beam dumping system of the FCC where triggering delay should be as short as possible. Another possible application is thyratron replacement in the PS injection generators, where the required switch dI/dt is above the capability of GTO-like thyristors triggered via their gate.

\section{Experimental setup}
\label{sec:setup}
\begin{figure} [htpb]
    \centering
    \includegraphics[width=0.6\textwidth,trim=0 0 0 0, clip,keepaspectratio]{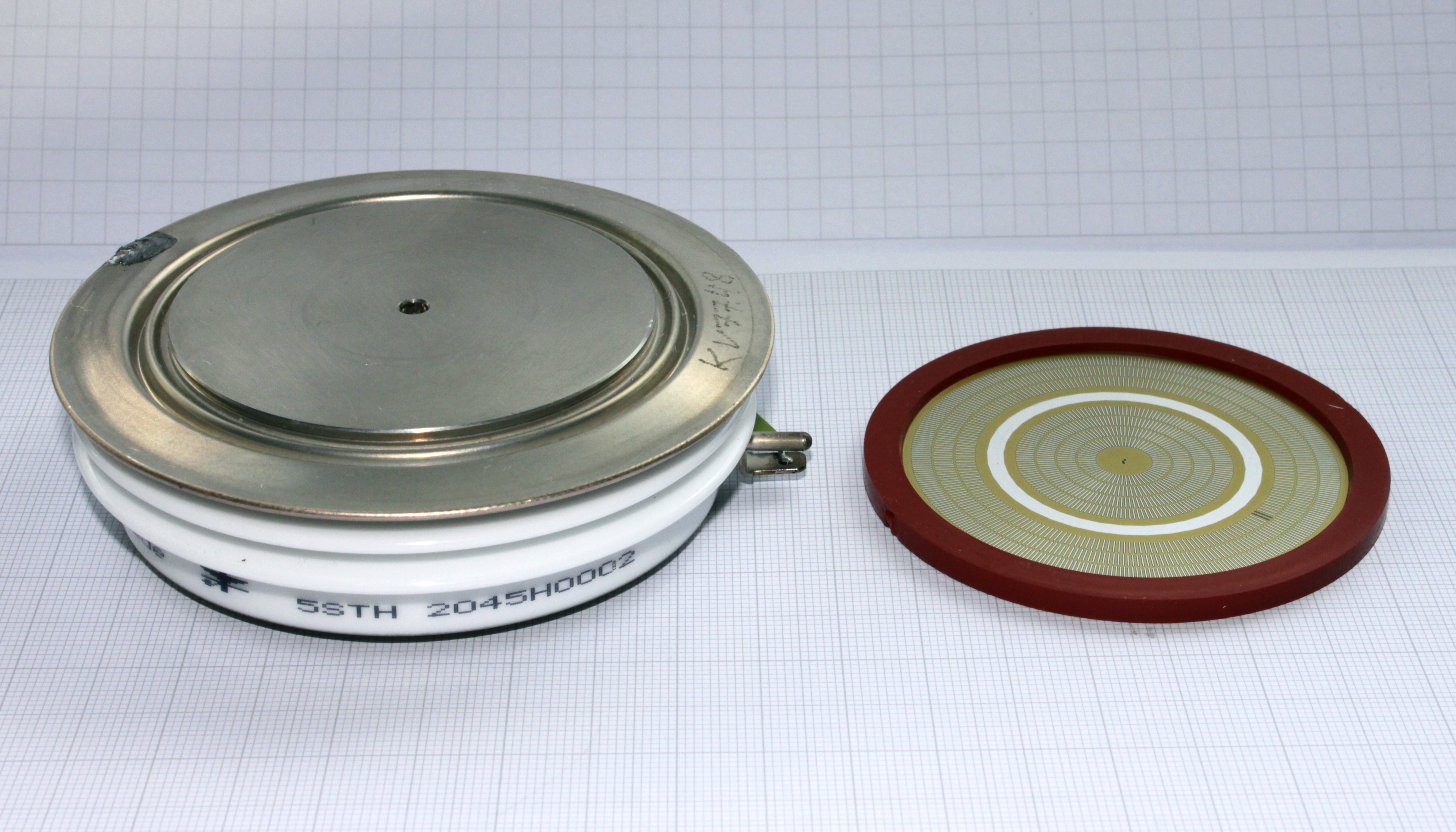}
    \caption{The appearance of the 5STH-20H450002 thyristor (left) and its semiconductor element (right).}
    \label{fig:Appearance}
\end{figure}

The GTO-like thyristor 5STH-2045H0002 presently used at CERN has been studied in impact-ionization triggering mode, both in a single pulse and in repetitive regimes. Its maximum repetitive peak forward blocking voltage VDRM is 4.5 kV (50 Hz, 10 ms), peak current pulse is 80 kA, and maximum rate of rise of current is 18 kA/\textmu s according to the datasheet. A more detailed description of this thyristor can be found in the datasheet \cite{Datasheet}. Photographs of the investigated thyristor and its wafer are shown in figure ~\ref{fig:Appearance}. The diameter of the silicon wafer is about 60 mm. A gate electrode in all experiments was connected to the cathode via a carbon ceramic resistor with a value of 1 $\Omega$.

\begin{figure} [htpb]
    \centering
    \includegraphics[width=0.7\textwidth,trim=0 15 0 15,clip,keepaspectratio]{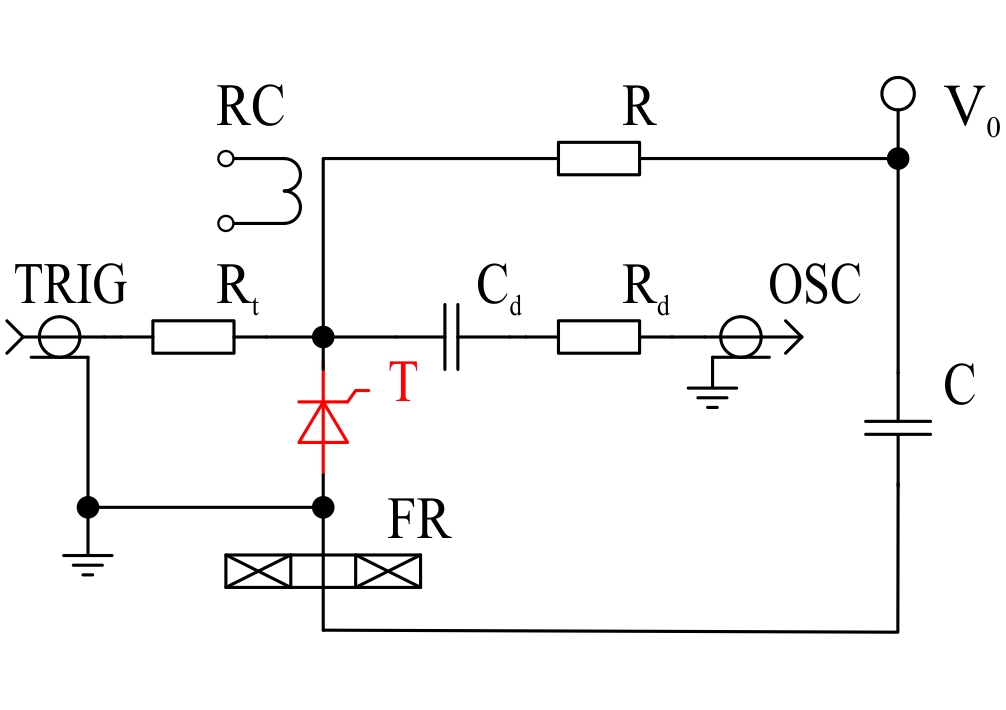}
    \caption{Circuit diagram of the experimental setup.}
    \label{fig:Circuit}
\end{figure}

The electrical circuit used in experiments is shown in figure~\ref{fig:Circuit}: it shows the experimental setup in general, although parts of the studies were conducted with modified circuit parameters. These changes are described below in further sections. The circuit consists of a capacitor bank C charged to negative polarity V\textsubscript{0} by either a DC or pulsed power supply (depending upon the test), the matched low-inductance load R, and the thyristor switch T triggered in impact-ionization wave mode by an external triggering generator TRIG.

A solid-state semiconductor opening switch (SOS) generator TRIG was used to trigger the \mbox{investigated} switch T, forming across 50-$\Omega$ matched load a negative-polarity pulse with the \mbox{amplitude} of up to 100 kV, a rise time of \verb'~'1 ns, and FWHM of \verb'~'4 ns. This trigger pulse is applied only across the T switch since the ferrite rings, FR, inductively decouple the remaining circuit elements. In addition, the power supply with charging voltage V\textsubscript{0} is connected via the charging inductor~L (not shown in figure~\ref{fig:Circuit}). Resistor R\textsubscript{t} allows to modify the rate of rise of the voltage applied across~thyristor~T. In the experiments, the value of resistance R\textsubscript{t} was varied from zero (resistor is short-circuited) up to~\verb'~'500 $\Omega$: the rate of rise of the triggering pulse across T is then in the range 6.0 to 0.8 kV/ns.

The Rogowski coil, RC, which has a rise time of 60 ns, was used to measure the current pulse through the investigated switch T. The voltage across the switch T was measured by the resistive divider described in \cite{Gusev_2018}. The upper arm of the divider is resistor R\textsubscript{d}~=~1 k$\Omega$, and the lower arm is formed by a coaxial cable with a characteristic impedance of 50~$\Omega$. The capacitor C\textsubscript{d} isolates the divider from the DC voltage V\textsubscript{0}. C\textsubscript{d} was short-circuited, in the experiments, when the capacitor C was charged in the pulsed mode. The intrinsic rise time of the voltage divider, determined during calibration, is \verb'~'250~ps between normalized amplitude levels of 10\% and 90\%. To measure the voltage across the switch T, a differential probe the Tektronix P5210 is also used. The measuring system includes a 30-GHz high voltage pulse attenuator produced by BARTH Electronics, a 12 GHz coaxial cable SFT-393 by Times Microwave Systems, and a 2.5-GHz digital real-time oscilloscope Tektronix DPO7254.

\begin{figure} [htpb]
    \centering
            \includegraphics[width=1.0\textwidth,trim=40 50 40 50,
            clip,keepaspectratio]{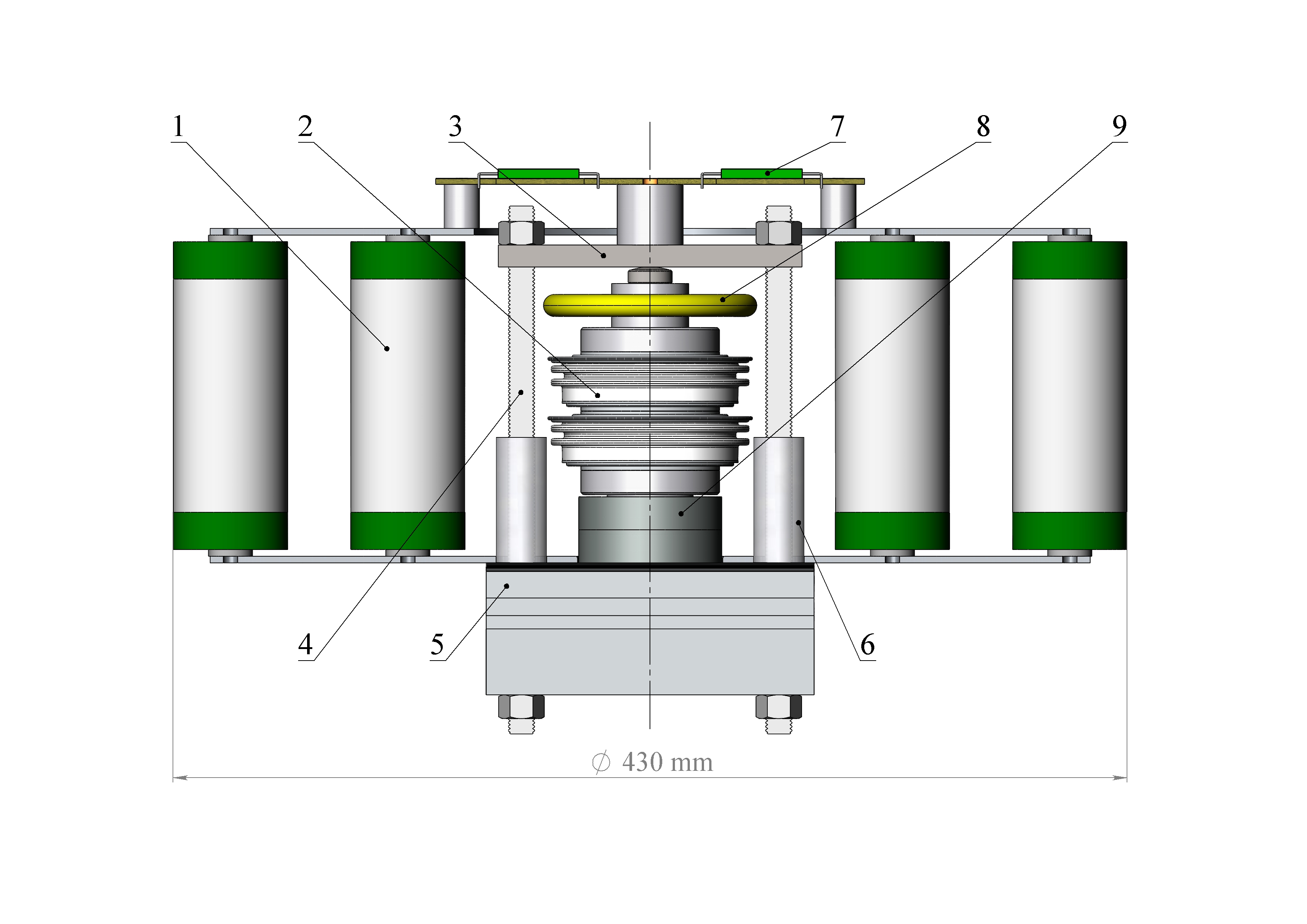}%
    \caption{The layout of the circuit elements: 1 --- capacitor bank, 2 --- thyristors, 3 --- crossbar, 4 --- metal rods, 5~---~heat sink, 6 --- dielectric sleeves, 7 --- resistive load, 8 --- Rogowski coil, 9 --- ferrite rings.}
    \label{fig:Layout}
\end{figure}

The simplified layout of the experimental circuit elements is shown in figure \ref{fig:Layout}. The discharging loop for capacitance C has a coaxial design. The capacitor bank C, which has an outer diameter of 430~mm, was assembled from film capacitors (item 1 in figure \ref{fig:Layout}) and had a total capacitance of 2~to~6~\textmu F, with an operating voltage of up to 10~kV. The thyristors under test (item 2) were located along the central axis of the setup.

The thyristors were mounted in a standard package, which includes a crossbar (item 3 in figure~\ref{fig:Layout}), metal rods (item 4), and a heat sink (item 5). The bushing (item 6) ensures adequate insulation between the high-voltage rods (item 4) and heat sink (item 5). As per the recommendations of the manufacturer of the thyristors \cite{Datasheet}, a mounting force of 22 kN was applied.

The resistive load (item 7 in figure~\ref{fig:Layout}) includes either 30 or 80 low-inductance carbon resistor connected in parallel: the nominal value of each resistor was 5.6~$\Omega$ or 51~$\Omega$ , respectively. The measured load resistance was 0.18~$\Omega$ or 0.63~$\Omega$ depending on the experimental setup. The Rogowski coil (item 8) measures the discharge current pulse. Voltage monitor (C\textsubscript{d}-R\textsubscript{d} in figure~\ref{fig:Circuit}, but not shown in figure~\ref{fig:Layout}) was connected as close as possible to the thyristors. The ferrite rings (item~9) K65x40x15~1000NN prevent the circuit influencing the triggering pulse from the SOS generator, but do not affect the discharge current pulse once the cores are saturated. The assembly has a total height of 300~mm.

The thyristor investigated has an asymmetric design and a maximum reverse voltage of 18 V. To prevent reverse voltage across the thyristors, a perfectly matched resistive load is required. The load value can be calculated from the capacitance and inductance of the discharge circuit. Capacitance could be measured, while the stray inductance is difficult to calculate or directly measure. Thus, a spark gap was installed in place of the thyristors to evaluate the stray inductance of the discharge circuit. Also, this allows one to determine the maximum discharge parameters, which can be obtained in the experimental setup.

Short circuit experiments were carried out at a charging voltage V\textsubscript{0} up to~10~kV, capacitance~C~=~6~\textmu F, and a short circuit in place of the load resistance (R~=~0 in figure \ref{fig:Circuit}). In addition, the thyristors were replaced by a spark gap. The period of the current oscillation was 4.5~\textmu s: hence, the calculated inductance of the discharge circuit is equal to 83~nH. This value was confirmed by a PSpice circuit simulation. Hence, the maximum dI/dt cannot exceed~120~kA/\textmu s at~V\textsubscript{0}~=~10~kV, for this geometry of experimental setup.

The value of load resistance has been optimized in PSpice circuit simulations. Maximum discharge current without significant reverse thyristor voltage (i.e.~$\ll$~18~V) was obtained with a load resistance in the range of 0.16--0.18 $\Omega$. Therefore, the load with resistance of 0.17 $\Omega$ was used to obtain the maximum possible discharge parameters in single pulse experiments; the resistance was changed to 0.16 $\Omega$ at the end of the experiments because of resistor degradation. In order to obtain the required discharge parameters, the resistance value of 0.63 $\Omega$ was chosen in the repetitive pulse experiments.

\section{Single pulse mode}
\label{sec:single}
\subsection{Switching stage}
\label{sec:switching}

\begin{figure} [htpb]
    \centering
    \setlength{\fboxsep}{-1pt}
            \includegraphics[width=0.7\textwidth,trim=0 5 0 10,
            clip,keepaspectratio]{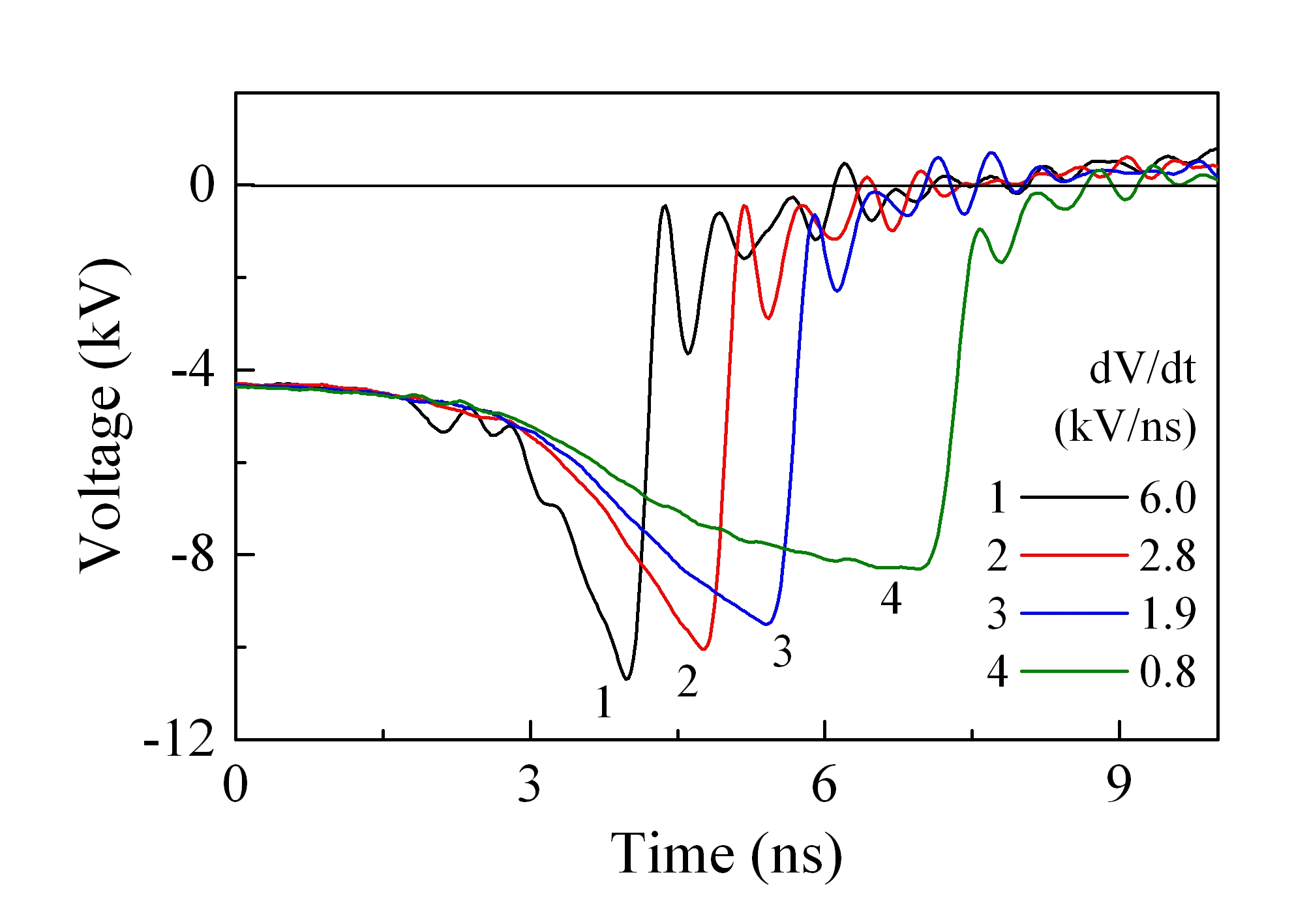}%
    \caption{Waveforms of the voltage across the thyristor during the switching process at different rates of rise of voltage of the triggering pulse.}
    \label{fig:TrigVoltage}
\end{figure}

First of all, the possibility of triggering the GTO-like thyristor in impact-ionization wave mode has been studied. A single thyristor was mounted as shown in figure \ref{fig:Layout}, the second thyristor in the stack was replaced by a short-circuit. Before the moment of triggering, a DC voltage of 4.2~kV was applied to the thyristor. Then, the thyristor was triggered as described in section \ref{sec:setup}. Increasing the resistance of the triggering resistor R\textsubscript{t} (figure \ref{fig:Circuit}) leads to a decreasing rate of rise of voltage. Measured voltage waveforms are shown in figure \ref{fig:TrigVoltage}.

The waveforms shown do not correct for the transient response of the voltage divider (\verb'~'250~ps). Also, the parasitic inductance and capacitance of the thyristor leads to ringing on the waveforms. Therefore, a switching time --- the duration of the voltage drop across the thyristor --- cannot be measured precisely. Nevertheless, it is seen in figure \ref{fig:TrigVoltage} that a fast sub-nanosecond switching occurred at all values of dV/dt in the range from 0.8~to~6.0~kV/ns: the switching time decreases with increasing dV/dt of the triggering pulse.

\begin{figure} [htpb]
    \centering
    \setlength{\fboxsep}{-1pt}
            \includegraphics[width=0.7\textwidth,trim=0 5 0 14,
            clip,keepaspectratio]{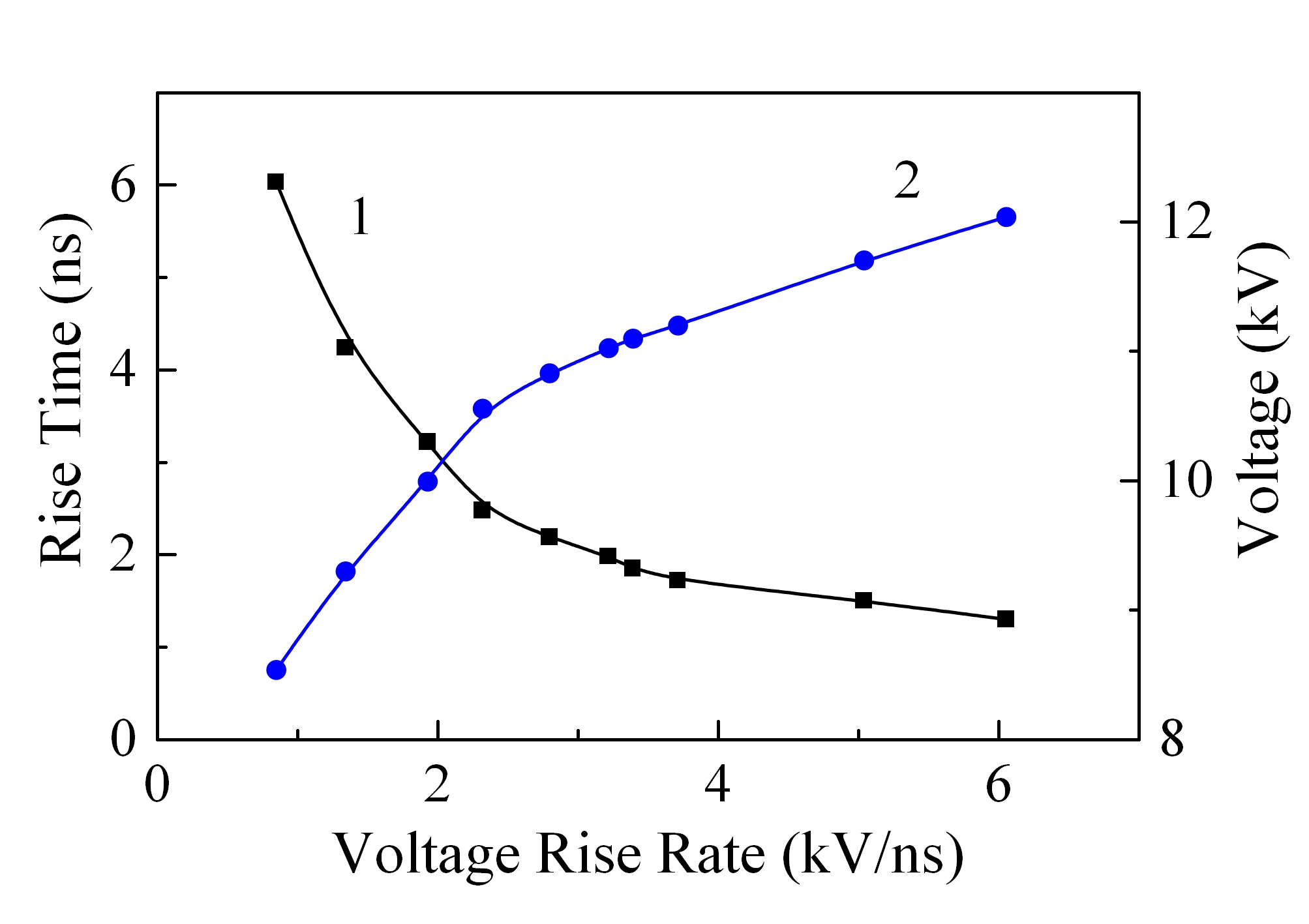}%
    \caption{The rise time of the triggering pulse (curve 1), measured 10\% to 90\%, and the maximum voltage across the thyristor (curve 2) as functions of the voltage rise rate.}
    \label{fig:Risetime}
\end{figure}

Figure \ref{fig:Risetime} shows the rise time of the triggering pulse and the maximum voltage across the thyristor V\textsubscript{m} as a function of dV/dt, in the range from 0.8~to~6.0~kV/ns: the transient response of the voltage divider was taken into account according to the technique described in \cite{Gusev_SC_2016}. The maximum voltage curve in figure~\ref{fig:Risetime} reflects the general feature of breakdown of all dielectric media, i.e., an increase in the breakdown voltage with a faster rate of rise of overvoltage pulse. In semiconductor devices, increasing voltage leads to a higher electric field in a p-n junction, which is the cause of the intensive charge carrier multiplication due to impact ionization. This explains the fast switching time at high dV/dt during the impact-ionization triggering.

\subsection{Current flow stage}
\label{sec:CurrentFlow}

In this part, we focus on the capacitor discharge process with maximum obtainable discharge parameters such as the current rate of rise, the maximum current amplitude, and the charging voltage. Two thyristors of type 5STH-2045H0002, connected in series, were assembled as illustrated in figure~\ref{fig:Layout}. Each thyristor has a circuit connected in parallel to ensure equal voltage distribution between the stacked thyristors (not shown in figure~\ref{fig:Layout}). These circuits consists of wire wound resistors and metal oxide varistors (MOV). Parasitic inductance of the wire wound resistor isolates the trigger wavefront from the MOVs on a short timescale. The capacitor bank includes 27~film capacitors connected in parallel. The measured capacitance of the assembly is 6~\textmu F, and the maximum operating voltage is 10~kV. Load resistance is 0.17~$\Omega$.

\begin{figure} [htpb]
    \centering
    \setlength{\fboxsep}{-1pt}
            \includegraphics[width=0.63\textwidth,trim=0 0 0 0,
            clip,keepaspectratio]{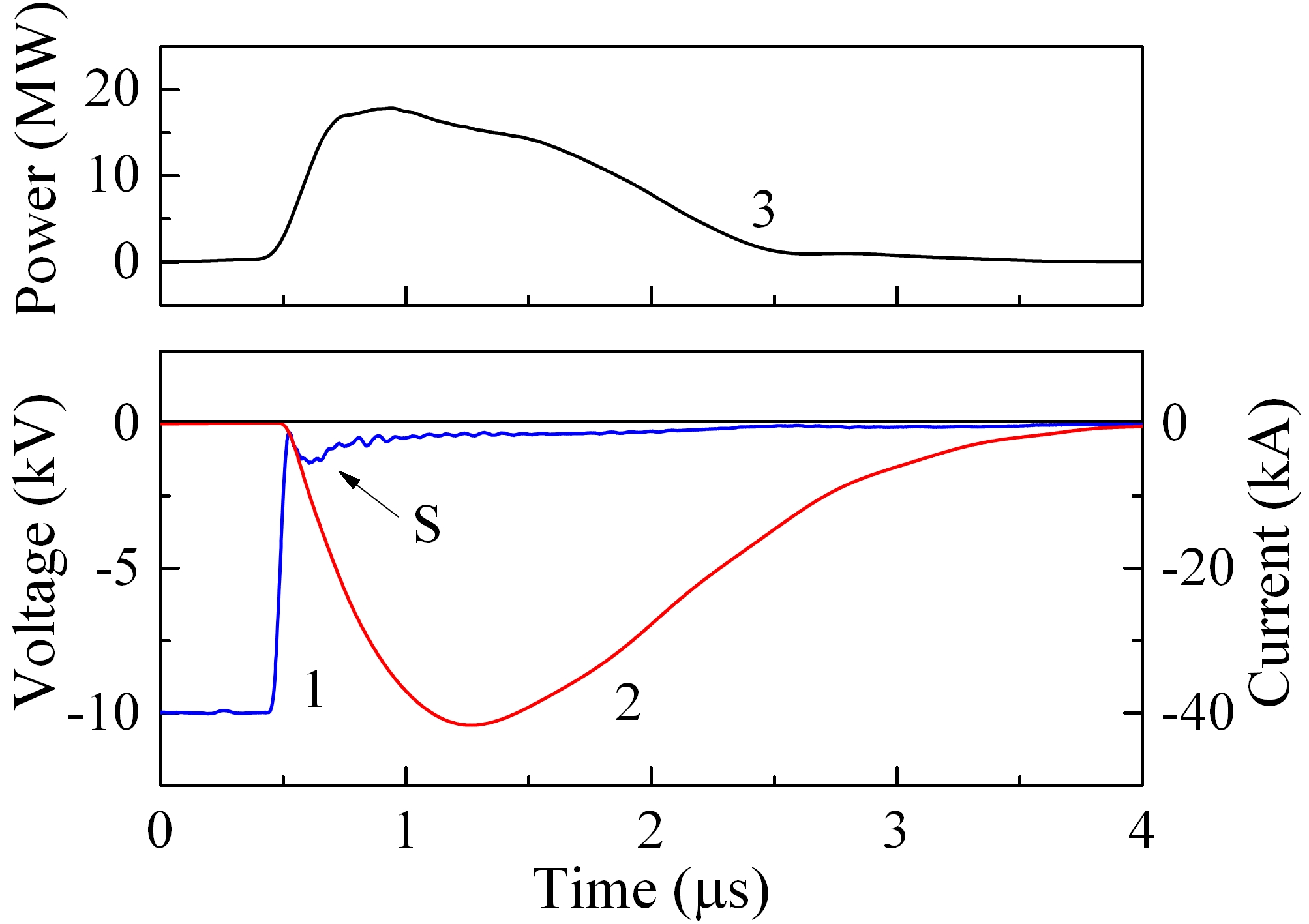}%
    \caption{Waveforms of the voltage (1), current (2), and power (3) across the stack of two series connected thyristors operated at the charging voltage of 10~kV.}
    \label{fig:10kV}
\end{figure}

Thyristors are switched into the conductive state by a forward direction overvoltage pulse from the SOS triggering generator. This pulse is applied to the two series connected  thyristors. Each thyristors gate electrode was connected to its cathode via a 1~$\Omega$ resistor all the time. The capacitor bank was charged to 10 kV by a DC power supply. The waveforms of voltage~(1) and current~(2) of the thyristor stack are shown in figure~\ref{fig:10kV}. The voltage is directly measured by the voltage divider connected as is shown in figure~\ref{fig:Circuit} (R\textsubscript{d}~=~10~k$\Omega$ and C\textsubscript{d}~=~0.44~\textmu F). This divider is used to measure a voltage in a microsecond range. The overvoltage trigger pulse, with duration of 2--3 ns, cannot be seen in curve~1, figure \ref{fig:10kV}, because of the intrinsic rise time of the voltage divider is equal to \verb'~'50~ns. The current waveform is calculated as an integral of dI/dt measured by the Rogowski coil. The power loss curve~3 is plotted by multiplying curves~1~and~2.

The parasitic inductance of the thyristor stack influences the measured voltage. Therefore, a voltage correction method was implemented to improve the measurement accuracy. The intrinsic inductance L\textsubscript{i} of the two series connected thyristors was estimated at the level of 10~nH, or 5~nH for a single thyristor. Inductive voltage drop V\textsubscript{L} is calculated as the product of dI/dt and L\textsubscript{i}. The corrected voltage, shown in figure~\ref{fig:10kV}, is calculated by subtraction V\textsubscript{L} from the measured voltage, and hence represents the resistive part of the voltage drop. 

The parameters obtained starting at a charging voltage V\textsubscript{0}~=~10~kV are as follows. The current amplitude --- 43~kA, and maximum value of the current rise rate --- 120~kA/\textmu s. The current pulse duration is \verb'~'1.5~\textmu s (FWHM). The energy stored in capacitor bank W\textsubscript{0}~=~306~J, the thyristor energy loss W\textsubscript{T}~=~24~J. The switching efficiency $\eta$ =~0.92~is calculated as
$\eta$~=~(1~--~W\textsubscript{T} / W\textsubscript{0}). The resistance of each thyristor R\textsubscript{Imax} $\approx$ 12 m$\Omega$ is determined at the moment of the maximum current.

The study of the influence of dV/dt, of the triggering pulse, on the thyristor at the current flow stage is the next important part of the work. The dV/dt was changed from 0.8~to~6.0~kV/ns, as was described in section \ref{sec:switching}. A single thyristor was investigated at a reduced value of stored energy of 18~J. A capacitor bank comprising of 9 film capacitors connected in parallel with a total capacitance of C~=~2~\textmu F was charged to V\textsubscript{0}~=~4.2~kV by the external pulsed power supply. The current pulse parameters are as follows: maximum current amplitude I\textsubscript{m}~=~5.5~kA, current rate of rise dI/dt~=~40~kA/\textmu s.

Attention was paid to the switching process to determine thyristor energy loss W\textsubscript{T} and switching efficiency $\eta$. The energy loss W\textsubscript{T} was calculated as an integral of the thyristor power loss (curve~3 in figure~\ref{fig:10kV}). The calculations took into account the parasitic inductance of the investigated thyristor. In addition, the resistance of the thyristor at different moments was estimated from Ohm's law. The first resistance value R\textsubscript{0} corresponds to the initial period of the current flow when the voltage spike across the thyristor is observed (S in figure~\ref{fig:10kV}). The second resistance value R\textsubscript{Imax} is calculated at the moment of the maximum of the discharge current flowing through the thyristor. Results are listed in table \ref{tab:I}.

\begin{table}[htbp]
    \centering
    
    \caption{Efficiency of the switching process for the thyristor. dV/dt~--- triggering voltage rate of rise, R\textsubscript{0}~--- initial resistance of the thyristor at time of the voltage spike, R\textsubscript{Imax}~--- resistance of the thyristor at a maximum value of the discharge current, W\textsubscript{T}~--- energy loss in the thyristor, W\textsubscript{0}~=~18~J~--- energy stored in the capacitor bank, and $\eta$ --- switching efficiency.}
    
    \smallskip

    \begin{tabular}{|c||c c c c c c c|}
        \hline
        dV/dt (kV/ns) & 6.0 & 5.0 & 3.7 & 2.8 & 1.9 & 1.3 & 0.8 \\
        \hline \hline
        R\textsubscript{0}~(m$\Omega$)  & 140 &	150 & 180 & 210 & 240 & 300 & 370 \\
        \hline
        R\textsubscript{Imax}~(m$\Omega$)  & 12 & 12 & 13 & 16 & 24 & 31 & 52 \\
        \hline
        W\textsubscript{T}~(J) & 0.29 & 0.30 & 0.35 & 0.43 & 0.56 & 0.72 & 1.13 \\
        \hline
        $\eta$~=~(1~--~W\textsubscript{T} / W\textsubscript{0})  & 0.98 & 0.98 & 0.98 & 0.97 & 0.97 & 0.96 & 0.94 \\
        \hline
    \end{tabular}
    
    \label{tab:I}
\end{table} 

\begin{figure} [htpb]
    \centering
    \setlength{\fboxsep}{-1pt}
            \includegraphics[width=0.7\textwidth,trim=0 5 0 14,
            clip,keepaspectratio]{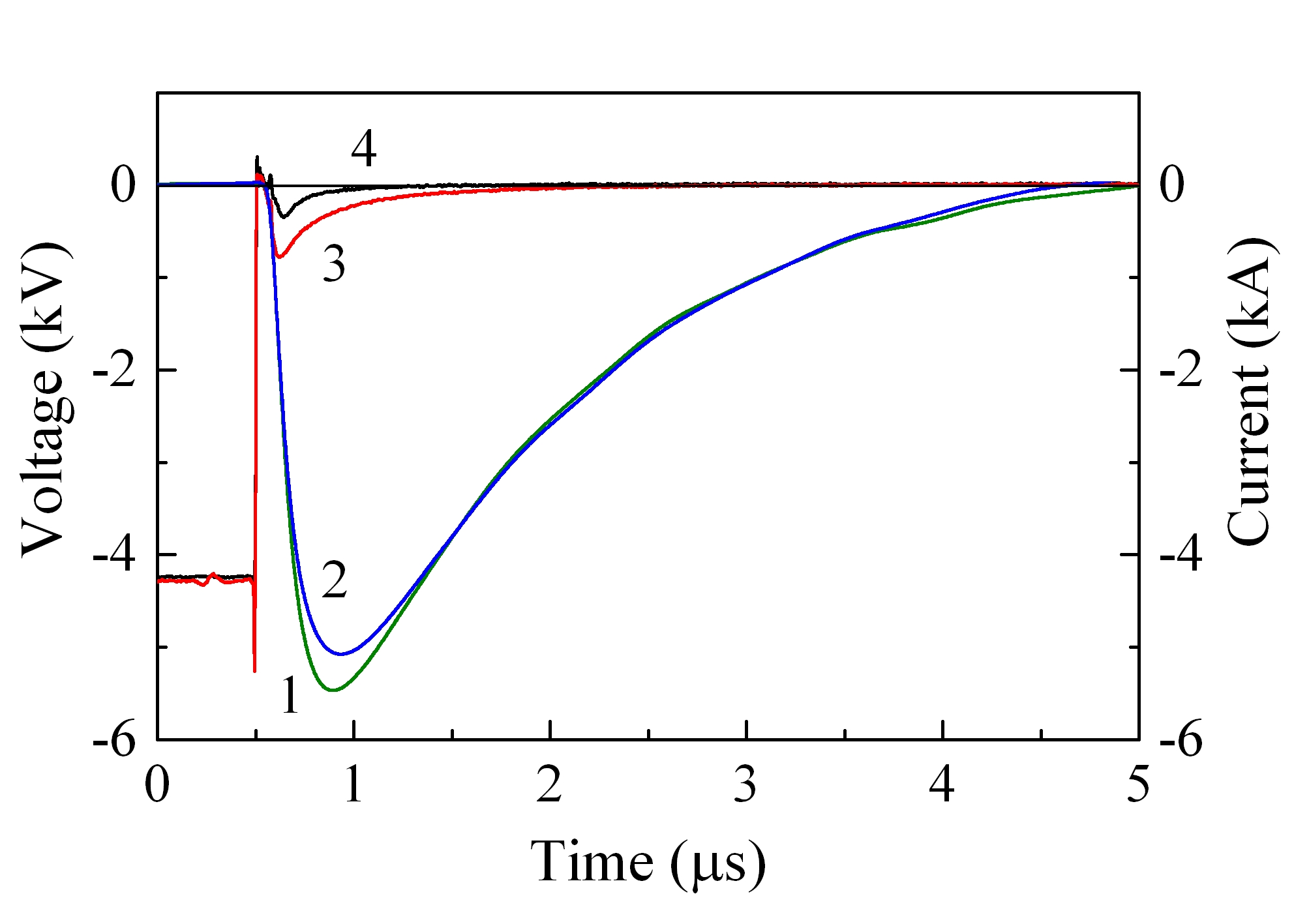}%
    \caption{Waveforms of the discharge current (1 and 2) flowing through the thyristor and voltage (3 and 4) across it at the triggering voltage rise rate 6~kV/ns (1 and 4) and 0.8~kV/ns (2 and 3).}
    \label{fig:Discharge}
\end{figure}

Typical waveforms of the discharge current and voltage across the thyristor at two extreme values of dV/dt (0.8 and 6.0~kV/ns) are illustrated in figure \ref{fig:Discharge}. While the currents (curves~1 and 2) have almost the same amplitude of \verb'~'5~kA, the forward voltage drop at the beginning of the current flow at dV/dt~=~0.8~kV/ns (curve~3) is twice as high as at dV/dt~=~6.0~kV/ns (curve~4). This effect has previously been observed and explained in conventional low dI/dt thyristors triggered in impact-ionization wave mode 
\cite{Gusev_2018}.

\begin{figure} [htpb]
    \centering
    \setlength{\fboxsep}{-1pt}
            
            \begin{multicols}{2}%
                        \includegraphics[width=0.5\textwidth,trim=0 175 0 8,      clip,keepaspectratio]{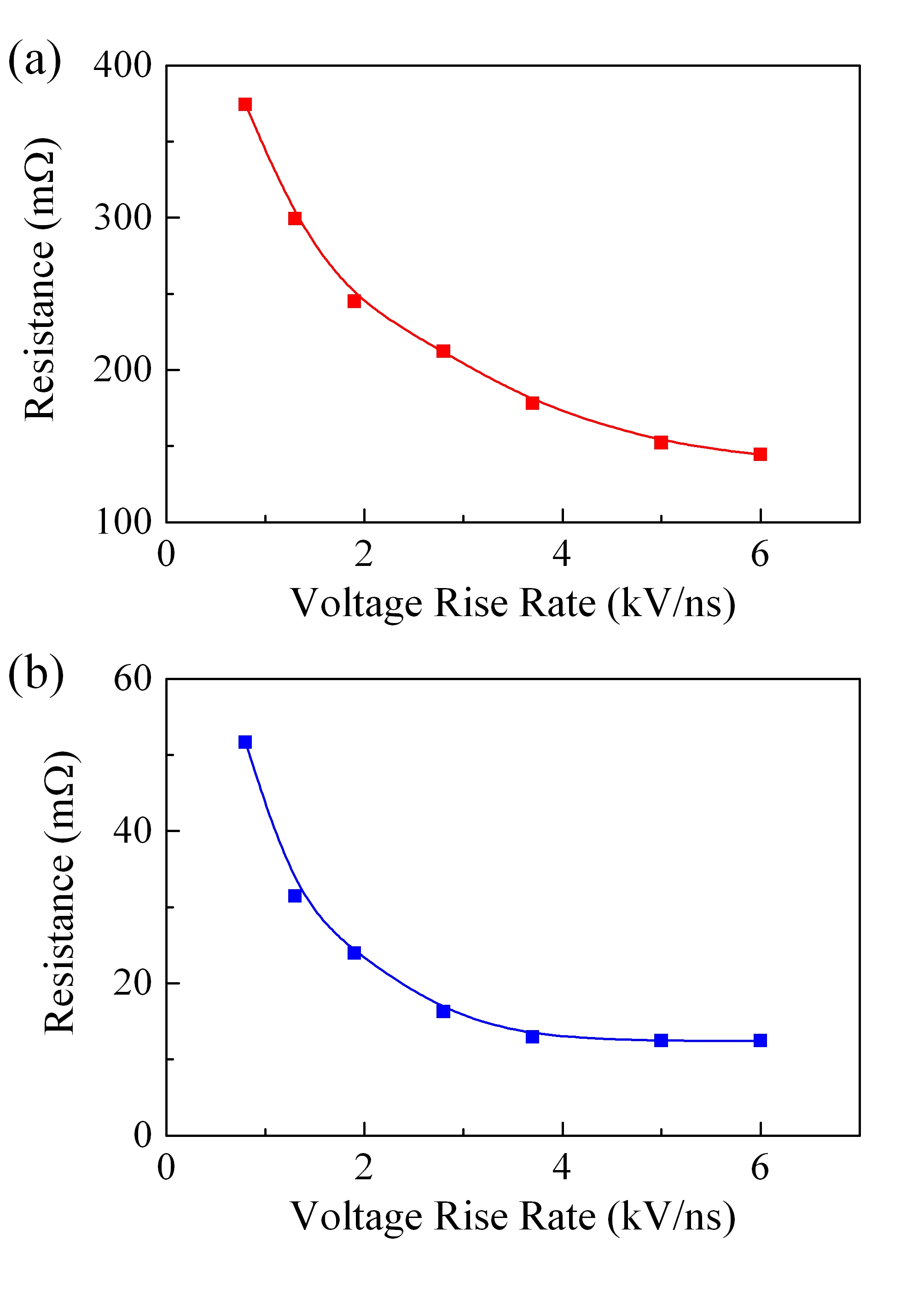}%
                        \hfill
                        \includegraphics[width=0.5\textwidth,trim=0 15 0 168,      clip,keepaspectratio]{Images/Fig_08.jpg}%
            \end{multicols}%
    \caption{The resistance of the thyristor as a function of the voltage rise rate at the different moments: (a)~--- at the time of the voltage spike in Figure~\ref{fig:Discharge}, (b) --- at the maximum of the discharge current.}
    \label{fig:Resistance}
\end{figure}

It is convenient to present some data from table \ref{tab:I} in the form of graphs. The resistance of the thyristor as a function of the voltage rate of rise at different moments is plotted in figure~\ref{fig:Resistance}. When current begins to flow through the thyristor after it triggers, the initial electron-hole plasma created by the impact-ionization wave is removed from the structure. Because of this, the voltage spike occurs (curves~3 and 4 in figure~\ref{fig:Discharge}). Therefore, the resistance at this moment is relatively high and is equal to hundreds of milliohms (figure~\ref{fig:Resistance},~a). After this, when the plasma concentration increases due to the current flowing through the injection junctions, the resistance mostly depends on the intrinsic structure of the thyristor. At the moment of maximum current, the resistance is equal to tens of milliohms (figure~\ref{fig:Resistance},~b). The resistance depends on the dV/dt of the triggering pulses in both cases. However, the shape of these dependencies differs. The curve in figure~\ref{fig:Resistance},~a shows that increasing dV/dt can decrease the resistance at the time of the voltage spike. On the contrary, the resistance at peak current is unchanged for dV/dt~>~4~kV/ns, as can be seen in figure~\ref{fig:Resistance},~b. Further numerical simulations are required to explain these dependencies.

\newpage

\begin{figure} [htpb]
    \centering
    \setlength{\fboxsep}{-1pt}
            \includegraphics[width=0.5\textwidth,trim=0 5 0 14,
            clip,keepaspectratio]{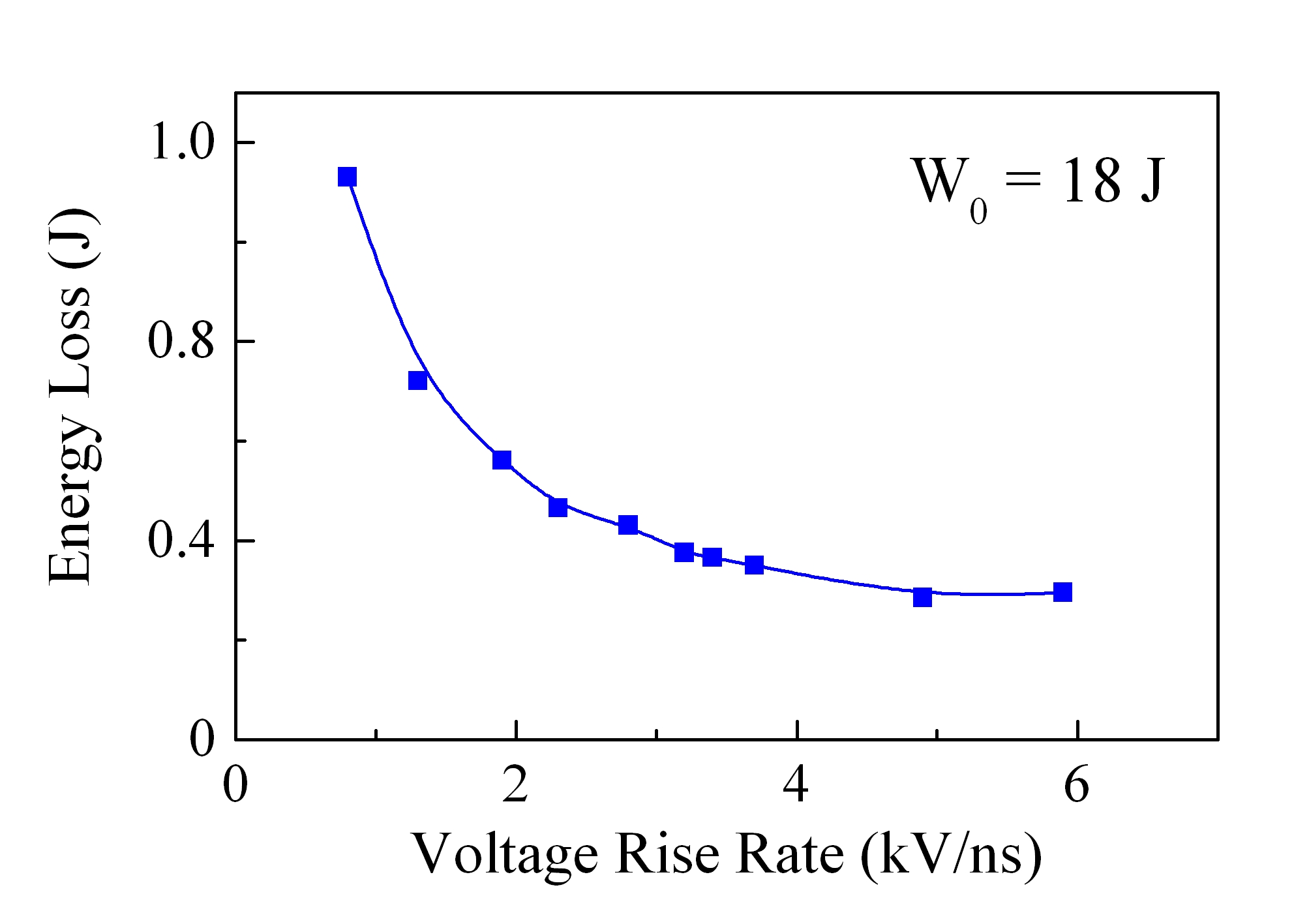}%
    \caption{Thyristor energy loss as a function of the voltage rise rate, for a total stored energy W\textsubscript{0}~=~18~J.}
    \label{fig:EnergyLoss}
\end{figure}

As shown in figure \ref{fig:EnergyLoss}, a decrease in the value of dV/dt of the triggering pulse, across the anode-cathode, leads to an increase in energy loss. The thyristor energy loss increases sharply for dV/dt of less than 2~kV/ns. Further decreasing of dV/dt can lead to thyristor damage. In practice, it is advisable to trigger thyristors at a dV/dt of at least~2--3 kV/ns to reduce the energy loss in thyristors.

\section{Repetitive mode}
\label{sec:Repetitive}

This chapter describes the essential issues of the practical application of thyristor switches triggered in impact ionization mode. These are the pulse repetition mode of operation and thyristor lifetime. The single thyristor 5STH-2045H0002 was studied at the triggering parameter of dV/dt~=~6.0~kV/ns, which is the maximum obtainable value in the experimental setup (figure~\ref{fig:Layout}) and corresponds to a resistor R\textsubscript{t}~=~0 (figure~\ref{fig:Circuit}). The capacitor bank C in figure~\ref{fig:Circuit} includes 9 film capacitors K75-90 (0.22~\textmu F~$\pm$~10~\%, 10~kV) connected in parallel; the measured total capacity is 2~\textmu F. At a charging voltage of 4.2~kV the stored energy is 18~J. With a resistive load of 0.63~$\Omega$ the discharge current pulse has a maximum amplitude of 5.5~kA, the rate of rise of current is 40~kA/\textmu s, and FWHM is~\verb'~'1.5~\textmu s. Typical waveforms of the voltage across the thyristor and the current flowing through it are pictured in figure~\ref{fig:Discharge} (curves~1 and 4).

\begin{figure} [htpb]
    \centering
    \setlength{\fboxsep}{-1pt}
            \includegraphics[width=0.53\textwidth,trim=0 5 0 14,
            clip,keepaspectratio]{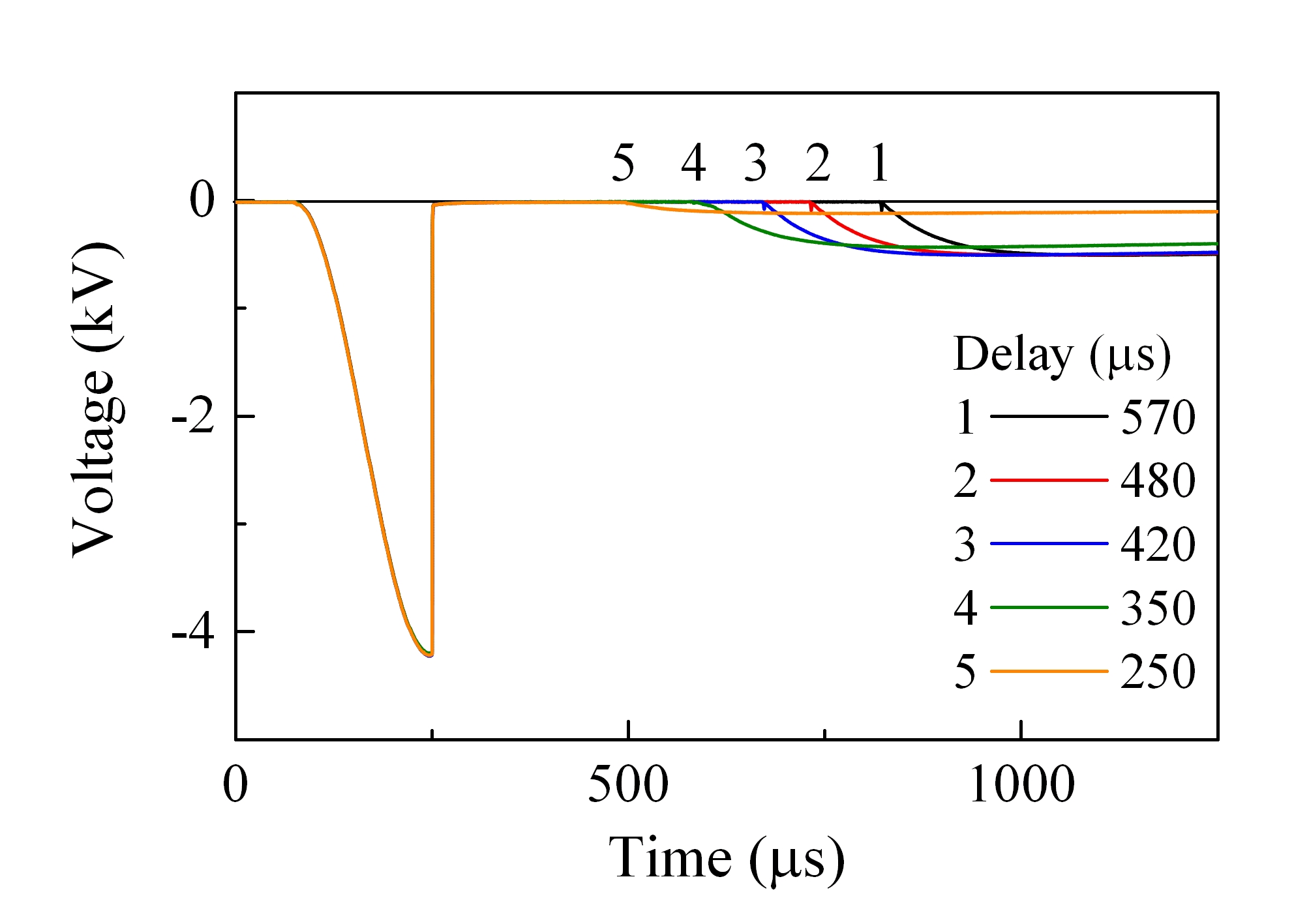}%
    \caption{Waveforms of the voltage across the thyristor at the charging and recovery stages.}
    \label{fig:Recovery}
\end{figure}

In order to estimate the maximum pulse repetition frequency (PRF) capability of the thyristor switch, in impact-ionization wave mode, the time of the thyristor recovery t\textsubscript{rec}, after passing high current, has been measured. Experiments were conducted with the above-mentioned discharge parameters and a charging voltage of 4.2~kV from the main power supply. An auxiliary power supply was also used, which consisted of a capacitor with a value of 1.3~\textmu F charged to \verb'~'1.3~kV. This capacitor was connected to the capacitor bank C via a charging inductor and an optically-controlled switch after the discharge cycle was complete. The capacitor bank charges up by 500~V after the optical-controlled switch is turned on. The time delay between the thyristor T triggering and the instant of turn-on of the auxiliary power supply for the next charging cycle was regulated from several milliseconds to tens of microseconds. The minimum time delay, for which the recovery time of the switch t\textsubscript{rec} was recorded, when the capacitor bank C began to charge, was 250~\textmu s as illustrated in figure \ref{fig:Recovery}.

However, it is necessary to discuss the thyristor recovery time in more detail. There is the delay time of more than \verb'~'400~\textmu s (curves~1, 2, 3 in figure \ref{fig:Recovery}) when the thyristor is stably recovered and the capacitor C is charged up to the calculated voltage of 500~V. The thyristor does not recover at all for a delay time of less than 250~\textmu s. When the charging voltage is applied with a delay time of 250--400~\textmu s, the thyristor is in an unstable state: the charging current from the second power supply partly flows through the thyristor. However, the thyristor recovers, but the capacitor bank C charges to a lower voltage (curves 4, 5 in figure \ref{fig:Recovery}).

\begin{figure} [htpb]
    \centering
    \setlength{\fboxsep}{-1pt}
            
            \begin{multicols}{2}%
                        \includegraphics[width=0.5\textwidth,trim=0 178 0 3,      clip,keepaspectratio]{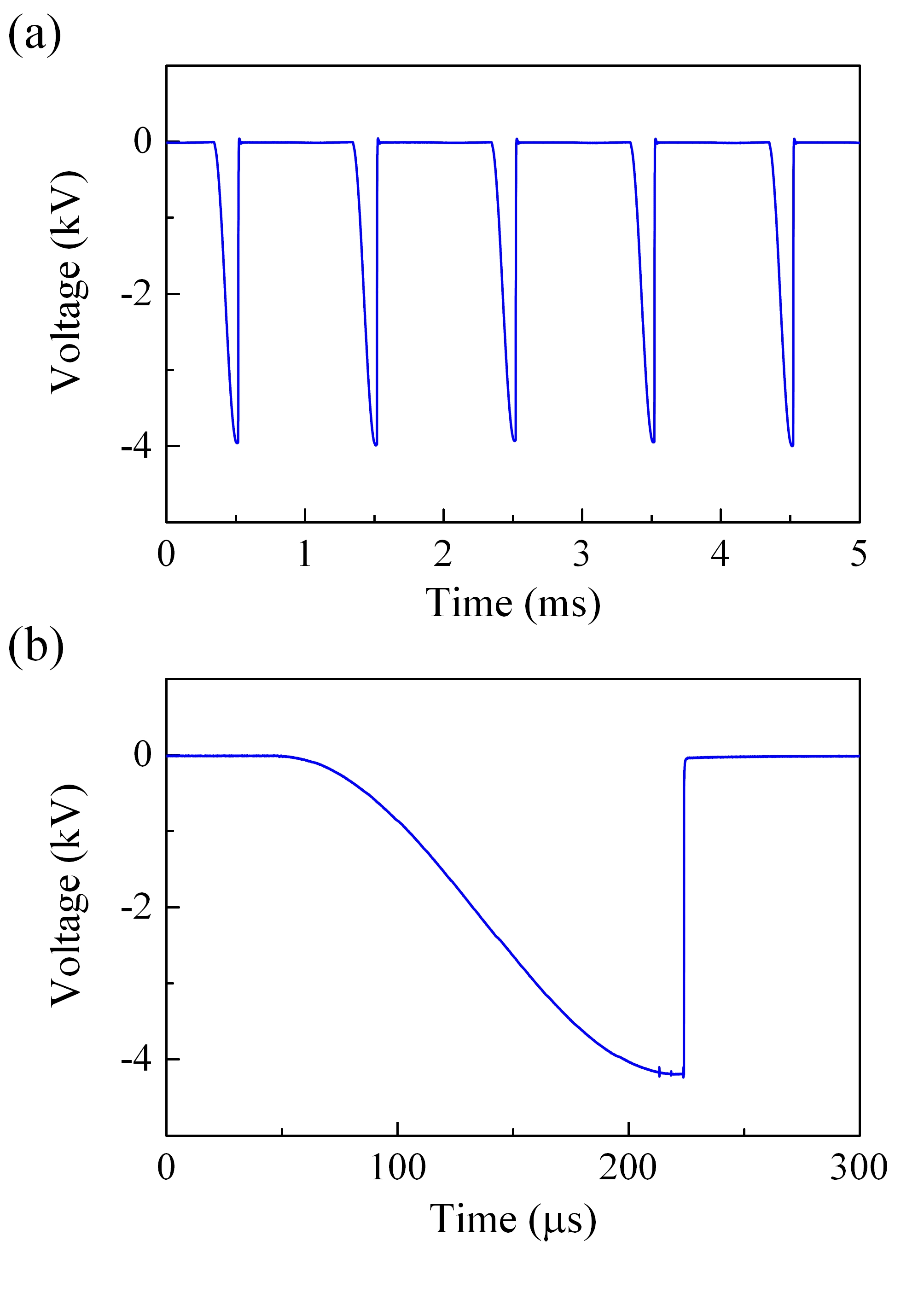}%
                        \hfill
                        \includegraphics[width=0.5\textwidth,trim=0 17 0 163,      clip,keepaspectratio]{Images/Fig_11.jpg}%
            \end{multicols}%
    \caption{Waveforms of the charging voltage across the single thyristor obtained in the burst operating mode at PRF~=~1~kHz: the pulse train (a) and the single pulse (b).}
    \label{fig:Repetitive}
\end{figure}

Taking into account the thyristor recovery time, the maximum PRF could be up to 4~kHz at t\textsubscript{rec}~=~250~\textmu s, and up to 2.5~kHz at t\textsubscript{rec}~=~400~\textmu s. However, in the experiments, the maximum PRF was restricted to 1~kHz by the external pulsed power supply. A single thyristor was tested in the repetitive pulse mode with discharge parameters as described at the beginning of this section, i.e. peak current of 5.5~kA. The voltage across the thyristor tested in repetitive mode is shown in figure~\ref{fig:Repetitive}. The nanosecond trigger pulse is not seen in microsecond timescale. In the continuous mode, the PRF was equal to 17~Hz; in the burst mode, the PRF was equal to 1~kHz for a burst duration of 0.5~s. Forced cooling of the thyristor and the load will be required to operate at higher PRF in continuous mode.

Finally, the lifetime of the investigated thyristor triggered in the impact-ionization wave mode was estimated. A leakage current I\textsubscript{leak} measured before the experiments for a new thyristor 5STH-2045H0002 was equal to 5~\textmu A at V\textsubscript{0}~=~4.5~kV. This parameter was chosen as an indication of the good condition of the thyristors. The thyristors are considered damaged if the leakage current I\textsubscript{leak} exceeds a value of several milliamperes. After experiments in single pulse mode at V\textsubscript{0}~=~10~kV, the leakage current increased to I\textsubscript{leak}~$\approx$~20~\textmu A for each thyristor. In the repetitive mode, at a charging voltage of 4.2~kV, more than 10\textsuperscript{6} pulses were performed at PRF up to 1~kHz in the burst mode, but mostly at PRF of 17~Hz in the continuous mode. Degradation of the thyristor, in terms of the I\textsubscript{leak} increasing, was not observed.

\section{Conclusion}

This work shows the possibility of triggering GTO-like thyristors in impact-ionization wave mode. As an example, the 5STH-2045H0002 thyristors were tested.  These thyristors have a hold-off voltage of 4.5~kV and can conduct up to 80~kA pulses with the current rate of rise of 18~kA/\textmu s \cite{Datasheet}, with conventional triggering via the gate control electrode. The implementation of the impact-ionization wave triggering improves the current rate of rise capability by more than six times compared to the datasheet value. Moreover, the switching time of the thyristors triggered in impact-ionization wave mode lies in the sub-nanosecond range. In this case, the maximum dI/dt achieved, in these experiments, is not limited by the thyristor itself but mostly by the electrical circuit. All these benefits are obtained by applying to the thyristors' main electrodes an overvoltage pulse. The amplitude of this pulse is twice as high as the static breakdown voltage, and the dV/dt is greater than~1~kV/ns. These conditions are required for the initiation of the impact-ionization wave in the reversely biased p-n junction in Si.

Two thyristors connected in series were tested in single pulse mode at a stored energy of \verb'~'300~J and a charging voltage amplitude of 10~kV. The maximum obtained discharge parameters are: peak current --- 43 kA, current rise rate --- 120~kA/\textmu s, and FWHM~--- 1.5~\textmu s.

Other experiments were conducted on a setup designed for repetitive pulse mode. In this setup, the single thyristor was tested with a stored energy of ~18~J and a charging voltage amplitude of 4.2~kV. The discharge parameters were: peak current~--- 5.5~kA, current rise rate~--- 40~kA/\textmu s, and FWHM~--- 1.5~\textmu s. Thyristor switching time to the conductive state does not exceed 300~ps in all experiments, both in single shot and repetitive modes.

The influence of the triggering voltage rise rate dV/dt on the characteristics of the thyristor was investigated. It is shown that an increase in the dV/dt parameter reduces the energy loss in the thyristor. The switching efficiency of 98 \% is obtained at the dV/dt of more than \verb'~'4~kV/ns. In practice, a dV/dt of at least 2--3~kV/ns should be chosen to reduce the energy loss in the thyristors.

The thyristor recovery time was found to be approximately 250~\textmu s, which corresponds to a maximum possible PRF of 4~kHz. This time mostly depends on the intrinsic process of the electron-hole plasma recombination. However, the current flowing through the thyristor can influence the recovery time: this needs to be additionally studied in further research. In this work, a maximum PRF of 1~kHz, limited by the power supply, was obtained. The investigated thyristor can efficiently operate at PRF up to 2.5~kHz according to results of the measurement of the thyristor recovery time.

More than 10\textsuperscript{6} high current discharge pulses were performed during this research. The thyristor leakage current at 4.5~kV was slightly increased from 5 to 20~\textmu A after the first pulses. However, this value remains within acceptable limits, and an additional increase of the leakage current was not observed.

A GTO-like thyristor triggered in impact-ionization wave mode meets the requirements for solid-state switches used in CERN and other particle acceleration facilities. Besides, it provides switching parameters that are not currently achievable by using existing semiconductor switches. A significant difference in switching performance of GTO-like thyristors compared to the inexpensive standard thyristors triggered in impact-ionization wave mode was not observed.


\acknowledgments
The work was supported by RFBR Grants Nos.~17-08-00406 and 18-08-01390, and by RAS Program Project No.~10. The study in part was carried out on the equipment of the Collective use centre at the Institute of Electrophysics, Ural Branch, Russian Academy of Sciences.



\end{document}